\documentclass[printer]{aa}
\usepackage{graphicx}
\usepackage{color}
\usepackage{amssymb}
\usepackage{amsmath}
\usepackage{lscape}
\usepackage{txfonts}
\usepackage{longtable}
\usepackage[round]{natbib}
\usepackage{tikz}
\usepackage[percent]{overpic}
\usepackage{footnote}
\usepackage[hyphens]{url}
\usepackage{hyperref}
\usetikzlibrary{positioning,shadows,arrows,trees,shapes,fit}
\begin{document}
   \title{Machine Learning Search for Gamma-Ray Burst Afterglows in Optical Surveys}
   \titlerunning{ML Search for Gamma-Ray Burst Afterglows}
   \author{Martin Topinka}

   \institute{Czech Technical University, Faculty of Electronic Engineering, Dept. of Radioelectronics, Technick\'{a} 2, Prague 6, Czech Republic\\
              \email{martin.topinka@gmail.com}
             }

   \date{\today; accepted XXXX}
   
% \abstract{}{}{}{}{}
% 5 {} token are mandatory

  \abstract
%context
   {Thanks to the advances in robotic telescopes, the time domain astronomy leads to a large number of transient events detected in images every night. Data mining and machine learning tools used for object classification are presented.}
%aims
   {The goal is to automatically classify transient events for both further follow-up by a larger telescope and for statistical studies of transient events. A special attention is given to the identification of gamma-ray burst afterglows.}
%methods
   {Machine learning techniques is used to identify GROND gamma-ray burst afterglow among the astrophysical objects present in the SDSS archival images based on the $g'-r'$, $r'-i'$ and $i'-z'$ colour indices.}
%results
   {The performance of the support vector machine, random forest and neural network algorithms is compared. A joint meta-classifier, built on top of the individual classifiers, can identify GRB afterglows with the overall accuracy of $\gtrsim 90\%$.}
   {}


   \maketitle
%
%________________________________________________________________
\section{Motivation}

\subsection{Boom of Time Domain Astronomy}
Thanks to the advances in robotic astronomy we have been entering the era of the boom of the time domain astronomy \citep{2014htu..conf..215D, 2011arXiv1110.4655D}. Instead of just taking photos of the sky, we shoot videos. Following the Moore's law, the data volumes is growing exponentially, doubling the size every 18 months.
Speaking of existing ground based surveys, or instruments that are about to be built in near future, just in the optical spectral window, such as SDSS \citep{1998AJ....116.3040G}, CRTS \citep{2009ApJ...696..870D}, PTF \citep{2009PASP..121.1334R}, Pan-Starrs \citep{2002SPIE.4836..154K}, LSST \citep{2007AAS...21113702I} they generate data streams up to the size of Petabytes.

For example, LSST that is planned to be in operation from 2020 will produce 30,000 GB per night - that is equal to the size of the entire SDSS. An estimated number of transients detected using differential imaging (subtracting a reference and an observed image of the same patch of the sky) reaches the order of magnitude of 1,000,000 transient alerts/night. 50\% of it is bogus alerts caused by CCD defects, random pixel fluctuations, airplanes etc \citep{2012AAS...21915605T}. The human attention time does not scale, therefore an automated real-time classification is essential.

Typically, the true nature of a transient can be told from longer series of observations and by obtaining a spectrum. However, observational time at a large telescope is expensive. This leads the need of setting priorities of each transient event based on scientific interest.

The primary classification must be fast enough because certain science requires rapid follow-up, otherwise the source fades beyond detectability or interesting features are gone.

\subsection{Transient Zoo}
Either resulting from a comparison with a catalog or from reference image subtraction, the detected transients may be of many different origins.

They can be software analysis artefacts or hardware defects, non-astronomical sources (airplanes) or just random pixel fluctuations statistically common for large chips.

For example, they can be results from periodic or quasi-periodic changes in brightness (transiting exoplanet, variable stars, binary systems, quasars, blazars), cataclysmic events (both supernova Type Ia and core-collapse explosions,  gamma-ray bursts, tidal disruptions of stars near a black hole), flaring episodes (novae, dwarf novae) and near Earth Objects (asteroids, comets). The common denominator of most of the transients is that they look very similar in a single observation.

In this analysis, the focus is given to identifying gamma-ray bursts in optical multi-band images. 

\subsection{Gamma-Ray Burst Afterglows}
A classical example of transients that requires fast reaction to understand the underlying physics are gamma-ray bursts (GRBs) with their afterglows. GRBs are the most powerful explosions in the Universe of yet unexplained origin, with the energy of up $10^{53}$ ergs released within seconds \citep{2004RvMP...76.1143P}. As the outflow interacts with the interstellar medium, the kinetic energy of the outflow is transformed into radiation in shock and produces an afterglow. The forward shock propagates into the interstellar medium, while reverse shock propagates backward to the outflow and can reveal physical parameters of the outflow itself. The combined properties of the reverse and forward shocks put constraints on the level of magnetisation of the outflow and therefore on the possible progenitors of GRBs. Due to a small width of an interacting shell, the reverse shock is observable only for a short time after the burst, therefore rapid follow-up of the most likely GRB afterglow candidates is crucial.

Due to the relativistic beaming effect, only radiation within a small pitch angle  $\theta_j \sim 1/\Gamma$ is visible to an observer, where $\Gamma$ is the bulk Lorentz factor of the outflow. As the outflow slows down, the peak energy of the radiation drops to the optical band and the $\theta_j$ increases. As a consequence, \textit{orphan afterglows} \citep{2002ApJ...579..699N}, the afterglows without a prompt $\gamma$-ray radiation detected, are expected to exist, however they would reach dimmer peak in brightness compared to the standard afterglows due to the relativistic Doppler effect for off-axis observers. Determining the rate of GRB orphan afterglows would directly affect the size of the jet opening angle and therefore GRB true rate and total energy of the explosion.

In this paper, I propose a mechanism how to distinguish GRB afterglows from other sources in an image and to improve existing searches for GRB afterglows in wide field images.

\section{Classification}
\subsection{Identifying GRB Afterglows}

Today's searches often define a GRB afterglow in a naive way as a new uncatalogued, typically decaying, source detected during or shortly after the prompt phase in the error-box of the GRB.

A more advanced definition of GRB afterglow is derived from the  physics of the forward shock and would cover the fact that an afterglow spectrum is typically composed of power-law segments, each with power-law evolution of break-points $F = \nu^{-\alpha} t^{-\beta}$, where typical observed values are $\alpha \sim 0.7$ and $\beta \sim 1.0$. The afterglow behaviour is often more complex than a single power-law due to an extra energy injection, clumpy interstellar medium, the presence of the reverse shock, geometry of the outflow and from peaks as the different spectral peaks are passing through a given energy window of the detector.

These signs makes the analytical classification close to impossible. Moreover, the set of rules above says nothing about the behaviour of other sources in the field of view. The stage is open for machine learning techniques to use.

\subsection{Machine Learning Classification}
Machine learning is a common envelope for the set of algorithms with tuneable parameters that improve their performance in classifying new data points based on the experience on previously seen data.
The main advantage of machine learning is that the target function, the function that classify the sources, may remain unknown. It is learnt from previously seen data within the limits from the chosen class of models. Clearly, if the target function were known before the analysis, no machine learning would be necessary and the function would be applied directly.

A typical learning processes involves grabbing features (observables and derived variables) from observations, splitting the data set in the train test and the test set. Then the learning algorithm is applied on the train data set and the result is validated on the test set not seen before during the training phase. At this step, meta parameters of the algorithm can be adjusted. Once the best model is found, the learnt procedure is applied on new arriving data.

Eventually, if the source is examined further and the nature of the source is proved or disproved, this feedback is extremely valuable to update the known data and to refine the algorithm.

\subsection{Feature Engineering}

The success of applying machine learning is heavily determined  by the input data. The minimal set of parameters that enables reasonable separation between GRB afterglows and the other sources results from a combination of data availability together with astrophysical intuition.

To achieve homogeneity of the features, the observations of 84 GRB afterglows detected by GROND \citep{2008PASP..120..405G} in the years 2010 -- 2015 are used. GROND is a set of near-infrared optical and infrared cameras mount on a 2.2~m telescope located in Chile. The actual GROND images with GRB observations are \textit{not} public. To mimic rapid response, the first observation report of a GRB afterglow or a GRB afterglow candidate are collected from published GCNs \citep{2000AIPC..526..731B} and dedicated publications \citep{2011A&A...526A.153K}, \citep{2012A&A...548A.101N}.

To simulate the population of other sources in the images within the field of view, archival data from the SDSS catalog are used. The catalogued objects within the $12' \times 12'$ of the field, centred around each GRB position, are taken. Stars and quasars are considered for the test, while not only the main sequence stars but any star-like sources that were not classified as galaxies, quasars or near Earth objects in the SDSS survey. This data set may also likely include possibly highly variable stars, novae etc.
The numbers of GRB afterglows, quasars and stars used in the analysis are shown in Table~\ref{table:numbers}.

\begin{table}[!ht]
\caption{Number of objects of each class in the dataset of sources used in the analysis.}
\label{table:numbers}
\smallskip
\begin{center}
{\small
\begin{tabular}{cc}  % l = left, c = centered
\hline
\noalign{\smallskip}
Source type & Number of sources \\
\hline
\noalign{\smallskip}
GRB & 84 \\
QSO & 22 \\
star & 328 \\
\hline\
\end{tabular}
}
\end{center}
\end{table}

SDSS provides data in Sloan filters u' g', r', i', z', while  GROND detected the majority of GRB afterglows in the g', r', i', z'. To achieve maximal completeness, the u' filter is omitted in the analysis.

\begin{figure}[ht!]
   \centering
   \includegraphics[width=0.95\linewidth]{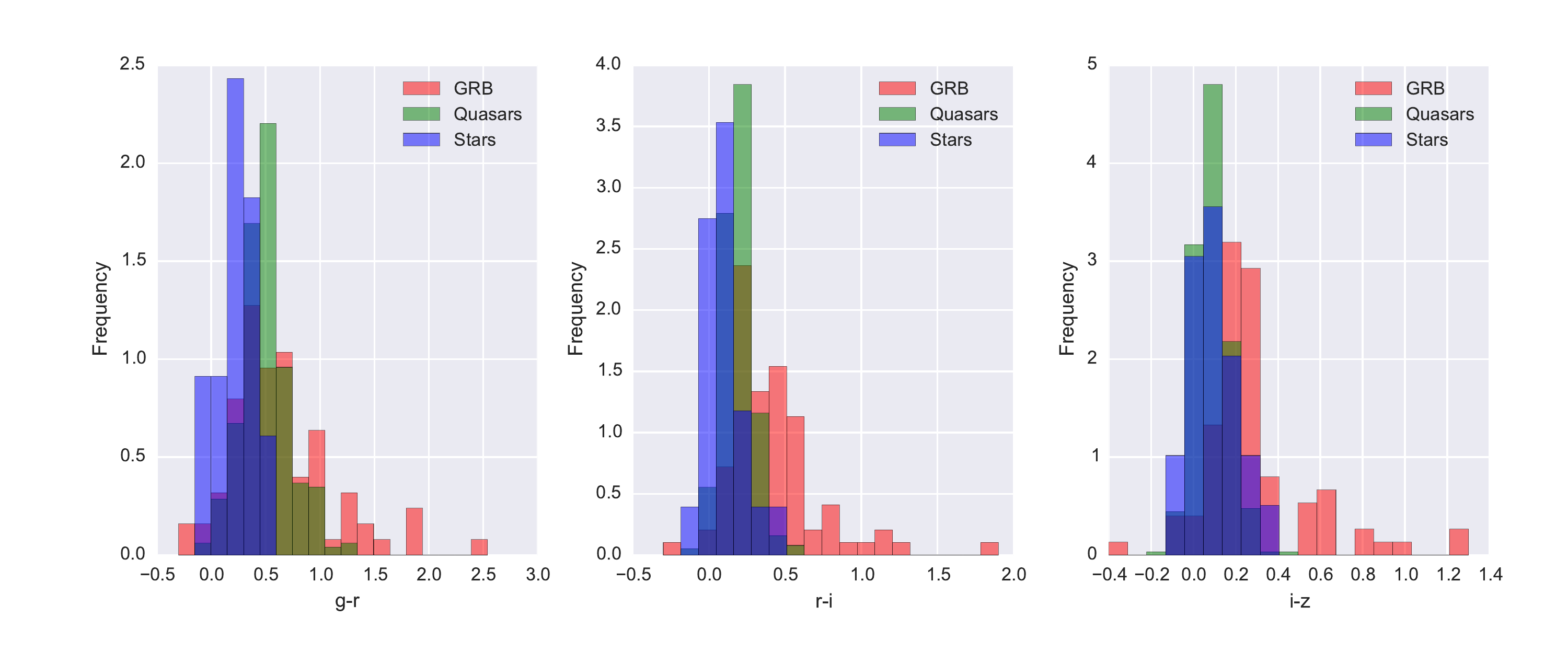}
   \caption{Histograms of g'-r' (left), r'-i' (middle) and i'-z' (right) colours based on different classes of objects: GRBs, quasars and stars.}
   \label{fig:color3hist}
\end{figure}

\begin{figure}[ht!]
   \centering
   \includegraphics[width=0.95\linewidth]{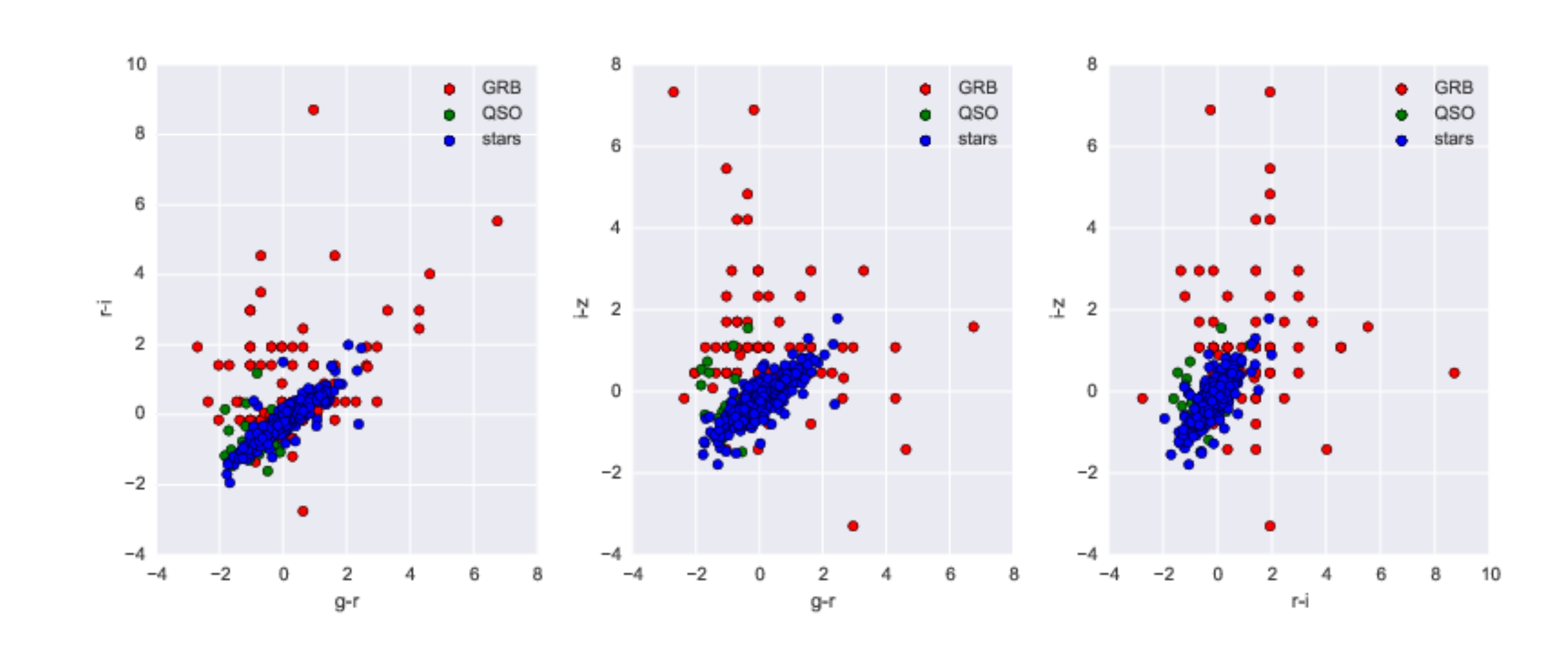}
   \caption{Colour-colour diagrams of relations between g'-r', r'-i' and i'-z' based on three different classes of objects: GRBs, quasars and stars.}
   \label{fig:color3}
\end{figure}

\begin{figure}[ht!]
   \centering
   \includegraphics[width=0.95\linewidth]{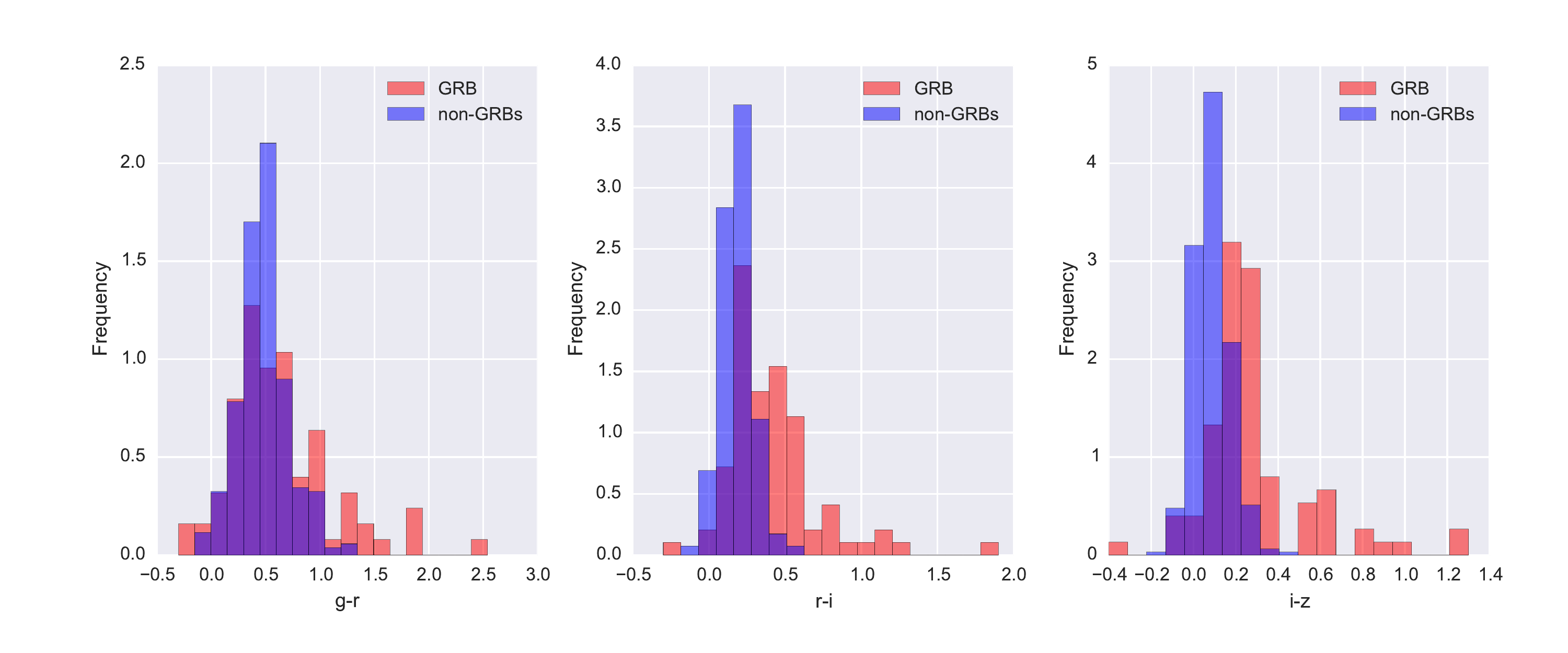}
   \caption{Histograms of g'-r' (left), r'-i' (middle) and i'-z' (right) colours visualised for GRBs and non-GRBs.}
   \label{fig:color2hist}
\end{figure}

\begin{figure}[ht!]
   \centering
   \includegraphics[width=0.95\linewidth]{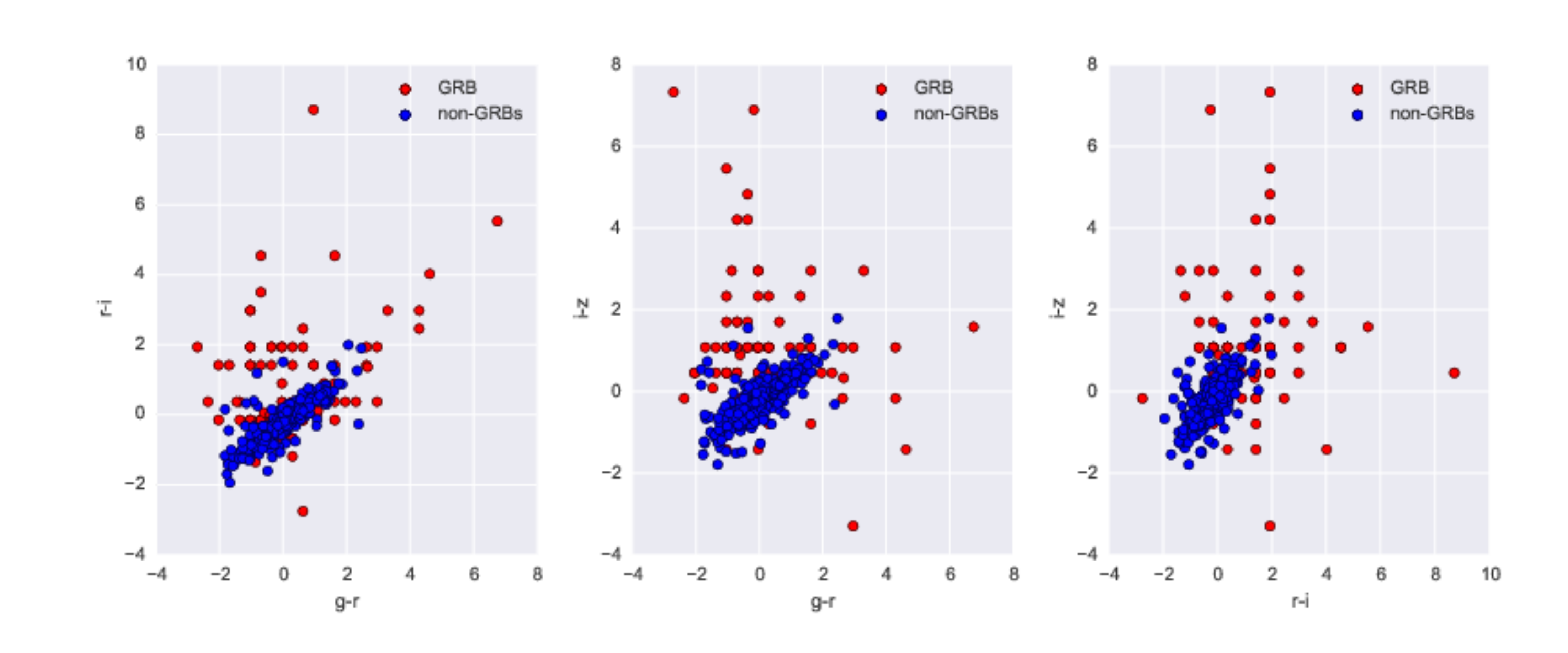}
   \caption{Colour-colour diagrams of relations between g'-r', r'-i' and i'-z' visualised for GRBs and non-GRBs.}
   \label{fig:color2}
\end{figure}

The distributions and scatter plots for each feature shows the over-lapping populations of GRB afterglows, quasars and stars (Fig.~\ref{fig:color3hist}, Fig.~\ref{fig:color3}) and the binary GRB vs non-GRB populations (Fig.~\ref{fig:color2hist}, Fig.~\ref{fig:color2}).

There are clearly patterns in the data, but the boundaries for delineating them are not obvious.
The degeneracy in each dimension does not allow to find a simple criterion to distinguish a GRB afterglow from other sources.

The temporal evolution of a GRB afterglow commonly follows a power-law decay in the light curve and characteristic, a power-law shape of a spectrum and often nearly constant, colour indices. The SDSS catalog provides several frames of each frame. Minimally, three time stamps would be necessary to check power-law behaviour. However, there is only a very small number of published multi-frequency afterglow light curves from GROND, thus temporal evolutionary data are not  used in the classification scheme here after.

At the early stage of detecting a transient, only small number of observations exists and only very little is known. Therefore, \textit{contextual information} may provide additional insight into the nature of the source. For example, a short distance to the nearest galaxy may indicate a supernova or a long GRB afterglow. The position in the Galactic plane sets higher priors for a galactic, stellar source.

The cross matching with other missions and with historical data listed in existing catalogs would highly improved the classification. The presence of the source counterpart at other wavelengths would be a strong lead, too. Typically, a transient localised in the error box of a GRB detected by a satellite would strongly favour a GRB. In the case of GROND, only one or two uncatalogued sources are detected in the GRB error box of reported GRBs, therefore, the positional context information is skipped in this analysis to prevent bias and a trivial solution. However, for surveys with large field of view, thinking of a transient as of a source that has changed brightness significantly, the contextual information would be essential for proper classification.

Summary of features potentially used to separate GRB afterglows from other transients is described in the Table~\ref{table:features}. Eventually, only $g'-r'$, $r'-i$ and $i'-'z'$ colour information is used for the analysis. It is a reasonable option, because both theoretical models and observational studies \citep{2004AIPC..727..487S} show that the colour indices of GRB afterglows does not vary significantly in time and therefore it diminishes the importance of an uncertainty at which phase of the afterglow was caught by the observation.  It also diminishes the scaling issues in observed magnitudes with no cosmological k-correction applied.

\begin{table*}[!ht]
\caption{Summary of features potentially used to separate GRB afterglows from other transients.}
\label{table:features}
\smallskip
\begin{center}
{\small
\begin{tabular}{llc}  % l = left, c = centered
\hline
\noalign{\smallskip}
Feature name & Description & Used in the classification\\
\noalign{\smallskip}
\hline
\noalign{\smallskip}
u', g', r', i', z' & magnitudes in 3 timestamps & yes, u' not used, 1 timestamp \\
u'-g', g'-r', r'-i', i'-z' & colour indices & yes, u'-g' not used \\
lcPL & deviation from PL light curve in $\chi^2$ & no \\
spePL & deviation from PL spectrum $\chi^2$ & no \\
$\sigma$ & variability defined as standard deviation & no \\
mad & absolute deviation from the median & no \\
 GRB & Inside a GRB error-box from a satellite? & no \\
X & Any counterpart in a the X-ray catalogue? & no \\
G& Any counterpart in a $\gamma$-ray catalogue? & no \\
O & Any counterpart in an optical catalogue? & no \\
galaxy & Normalised distance to the nearest galaxy & no \\
\noalign{\smallskip}
\hline\
\end{tabular}
}
\end{center}
\end{table*}

\subsection{Classifiers}

There is a large variety of different classifiers used in machine learning, each suitable for slightly different sort of problems. Sneaking the available data, over-lapping distributions of feature values and the pair plots of the features suggests that the problem is non-linear. Performance of three algorithms is tested: support vector machine classifier, random forest classifier and neural network classifier.

\subsubsection{Support Vector Machine}
Support vector machine \citep{Smola:2004:TSV:1011935.1011939} constructs a hyperplane in the feature space trying to maximise the distance (margin) to the nearest training-data point of any class. If linear separation is not possible kernel transformation is applied on data, adding an extra dimension, e.g. radial base function  $k(\mathbf{x_i},\mathbf{x_j})=\exp(-\gamma \|\mathbf{x_i} - \mathbf{x_j}\|^2)$, for $\gamma > 0.$. Thus, the separation is found in the hyper-space and projected back to the original parameter space.

\subsubsection{Random Forest}
Decision trees is a popular method for various machine learning tasks based on binary splitting the data space or its sub-space to achieve the best separation of data points belonging to different classes and therefore the maximal information gain. Unfortunately, decision trees often suffer from the risk of over-fitting. However, if averaging over a large ensemble of multiple decision trees (forest) is used, each of them with imputed imperfection, e.g. an omitted parameter, pruned tree branch, a more robust classifier is built \citep{Breiman:2001:RF:570181.570182}.

%Hellinger Distance Trees for Imbalanced Streams
Classification of rare events, such are GRB afterglows, typically suffers from poor generalisation performance if standard entropy or Gini splitting criteria are used. The Hellinger distance measure
\citep{Lyon:2014:HDT:2703763.2704125,Cieslak:2012:HDD:2124701.2124730} is used.

Random forest is also used to measure the relative importance of each feature for the classification task. The average classifying efficiency is measured over many random decision trees leaving one feature aside. The inverse of classification accuracy over a pruned tree defines the importance of the feature.

\subsubsection{Neural Network}
Artificial neural network is a composition of neurones/perceptrons. Each linear perceptron returns the weighted sum of inputs,  $f (x) = K \left(\sum_i w_i g_i(x)\right)$ , where $K$ is the activation function. Higher number of perceptrons combined together in layers can describe even highly complex non-linear behaviour \citep{Bishop:1995:NNP:525960}.

\section{Results}
\subsection{Validation}
To prevent memorising data points with only negligible generalisation rather than the ability to learn, it is important to test the performance of the algorithm on an independent data set to the data used for learning. Both the train and test sets should have similar statistical properties including similar abundance of observations within each class. To overcome the high variance of testing accuracy, the K-folding cross-validation method is used to randomly shuffle the data and to split the entire data set in 4 sub-groups, 3 of 4 sub-groups (75\%) of data is used to train the algorithm, while the remaining part (25\%) is used to test the algorithm, the same approach is applied repetitively with different 25\% of the data to be the test set. This way, each observation is used both in the train and test set. The overall performance is the average of individual accuracy scores. The folds are constructed with preserving the class abundances in each sub-samples.

To test the classification performance of each algorithm and to adjust meta parameters of each procedure, the confusion matrix (CM) and Receiver Operating Characteristics (ROC) curve are used.

\subsubsection{Confusion matrix}
The confusion matrix represents the numbers of true positive  ($t_p$) detections and true negatives ($t_n$) on the matrix diagonal and the numbers of false positives ($f_p$) and false negatives ($f_n$) on the anti-diagonal. Normalised form of CM is often used.

Following scores are constructed from $t_p$, $t_n$, $f_p$ and $f_n$ to measure the performance: $accuracy \equiv (t_p + t_n)/(t_p+t_n + f_p + f_n)$, $precision \equiv t_p/(t_p + F_p)$, $recall \equiv t_p / (t_p + f_n)$ and $F_1 \equiv 2~precision \times recall / (precision + recall)$. All range from 0 to 1, with 1 being the best.

GRB afterglows are rare events. If the classes are of very different sizes the overall classification accuracy is often not the best choice of the performance metrics, because the majority class likelihood prior is dominant. There are several option to fix it: a) to down sample randomly the non-GRB sample b) perform two-step classification, apply a simple binary classifier select the most obvious cases in the majority class, then remove them from the dataset and redo the classification with a new more balanced dataset c) use metrics that is more accurate for imbalanced classes, e.g. the Matthews correlation. The Matthews correlation coefficient (MCC)

\begin{equation}
\label{eq:metrics}
MCC = \frac{t_p \times t_n - f_p \times f_n}{\sqrt{(t_p + f_p)(t_p + f_n)(t_n + f_p)(t_n+f_n)}}
\end{equation}

is used in the search for the best hyper parameters of the classifying algorithms in this analysis.  MCC takes into account true and false positives and negatives \citep{journals/bioinformatics/BaldiBCAN00}.

\subsubsection{ROC curve}
The Receiver Operating Characteristics (ROC curve) is a parametric plot of the true positive rate against the false positive rate with a classification probability threshold as the parameter. It serves as a metrics that measures the classification performance of a binary classifier. The predictions to be positive or negatives are considered to be often over-lapping statistical variable. Typically, there is a trade off between the sensitivity of the classifier (characterised by the true positive rate) and a complement to the specificity reflecting false triggers (the false positive rate).

Any increase in sensitivity will be accompanied by a decrease in specificity. The closer the curve follows the left-hand border and then the top border of the ROC space, the more accurate the test, bearing in mind that the diagonal would represent the algorithm that classifies transients randomly. The area under the ROC curve is a measure of classification accuracy, especially for imbalanced classes.

In the case of the transient follow-up, low sensitivity would lead to missed detections, while low specificity would result in wasting of available follow-up resources, since there would be a high number of non-GRBs.

\subsection{Classifier Comparison}
Python \textit{scikit-learn} machine learning library \citep{Pedregosa:2011:SML:1953048.2078195} was used with customised implementation of neural network using multi-layer perceptron algorithm.

The results for Support Vector Machine (SVM), Random Forest (RF) and Neural Network (NN) classifiers are shown and summarised in the Table~\ref{table:accuracy}.

\begin{figure}[ht!]
   \centering
   \includegraphics[width=0.95\linewidth]{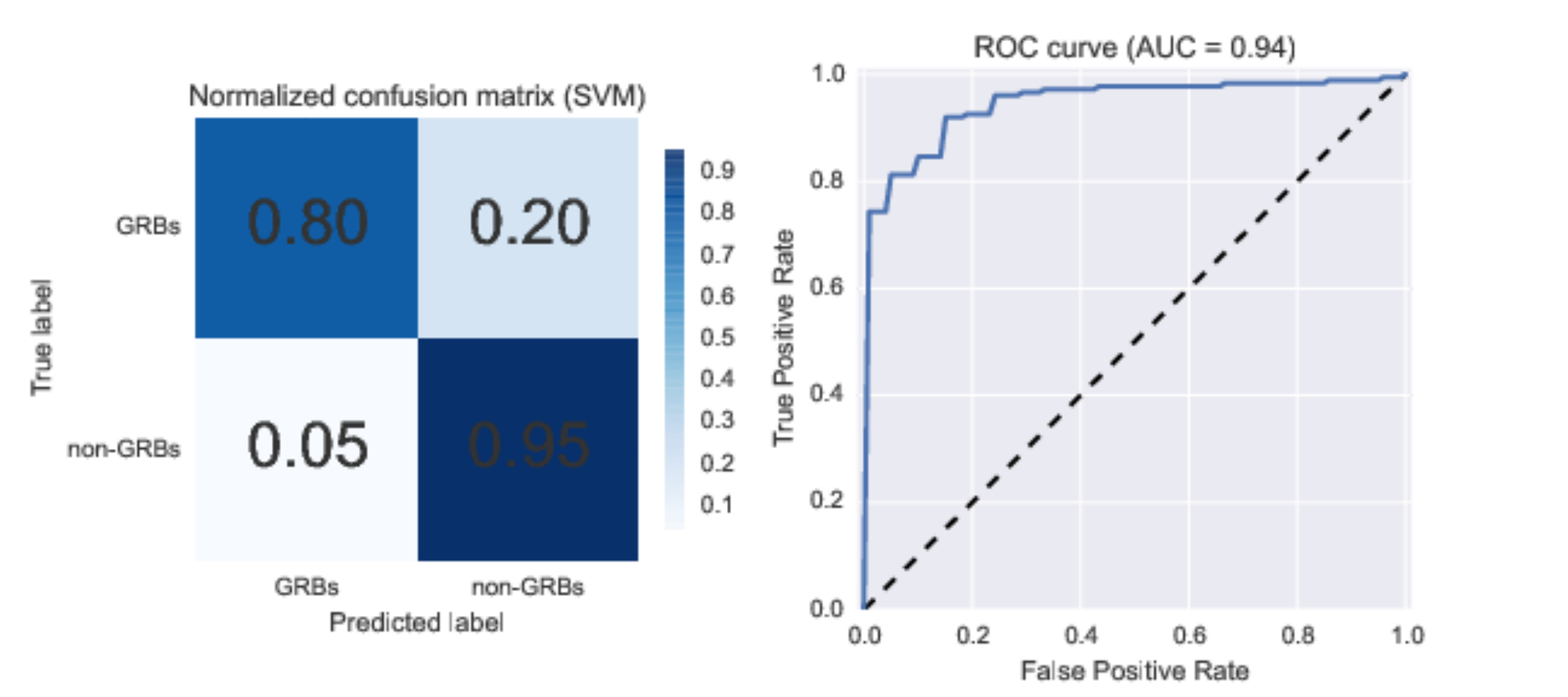}
   \caption{Confusion matrix (left) and the ROC curve (right) for the SVM-classifier. The overall accuracy is 0.90.}
    \label{fig:svm-cmroc}
\end{figure}

\begin{figure}[ht!]
   \centering
   \includegraphics[width=0.95\linewidth]{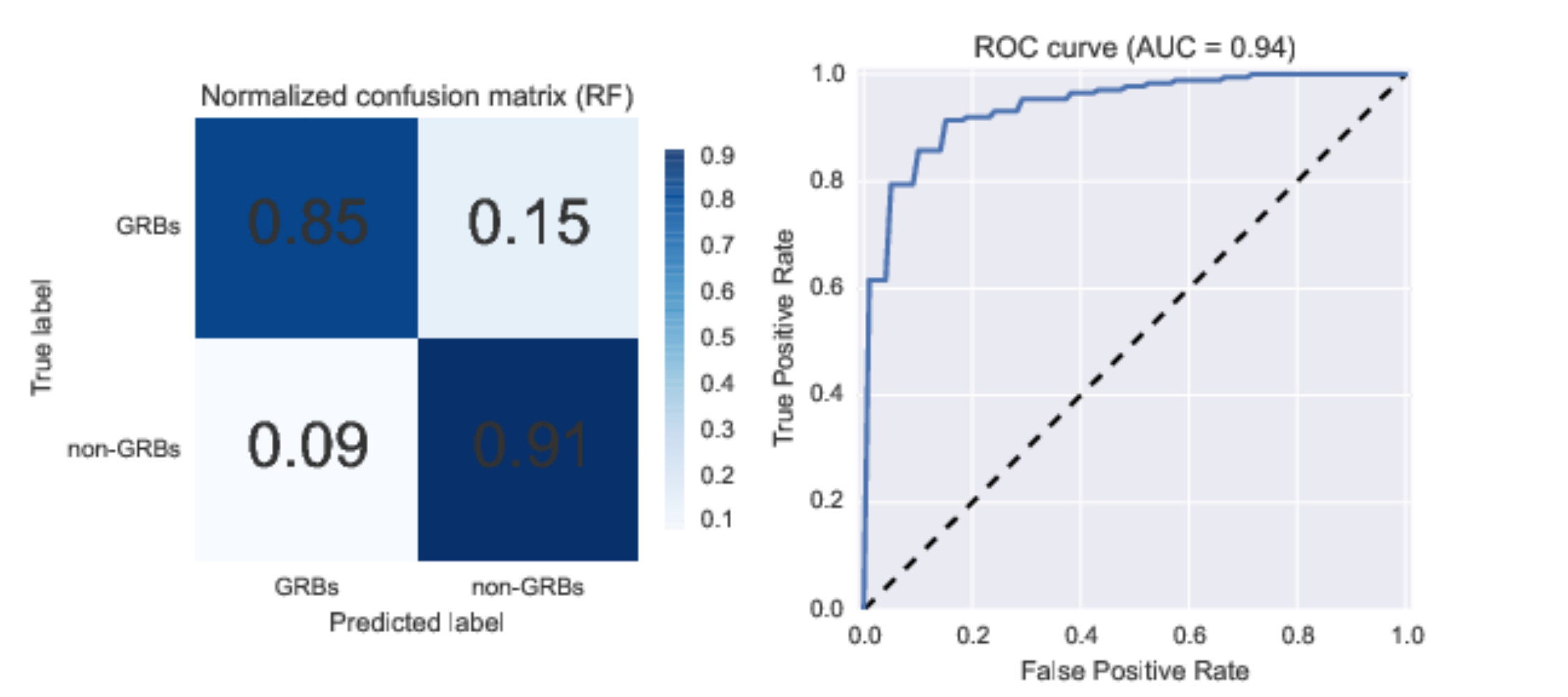}
   \caption{Confusion matrix (left) and the ROC curve (right) for the RF-classifier. The overall accuracy is 0.89.}
   \label{fig:rf-cmroc}
\end{figure}

\begin{figure}[ht!]
   \centering
   \includegraphics[width=0.95\linewidth]{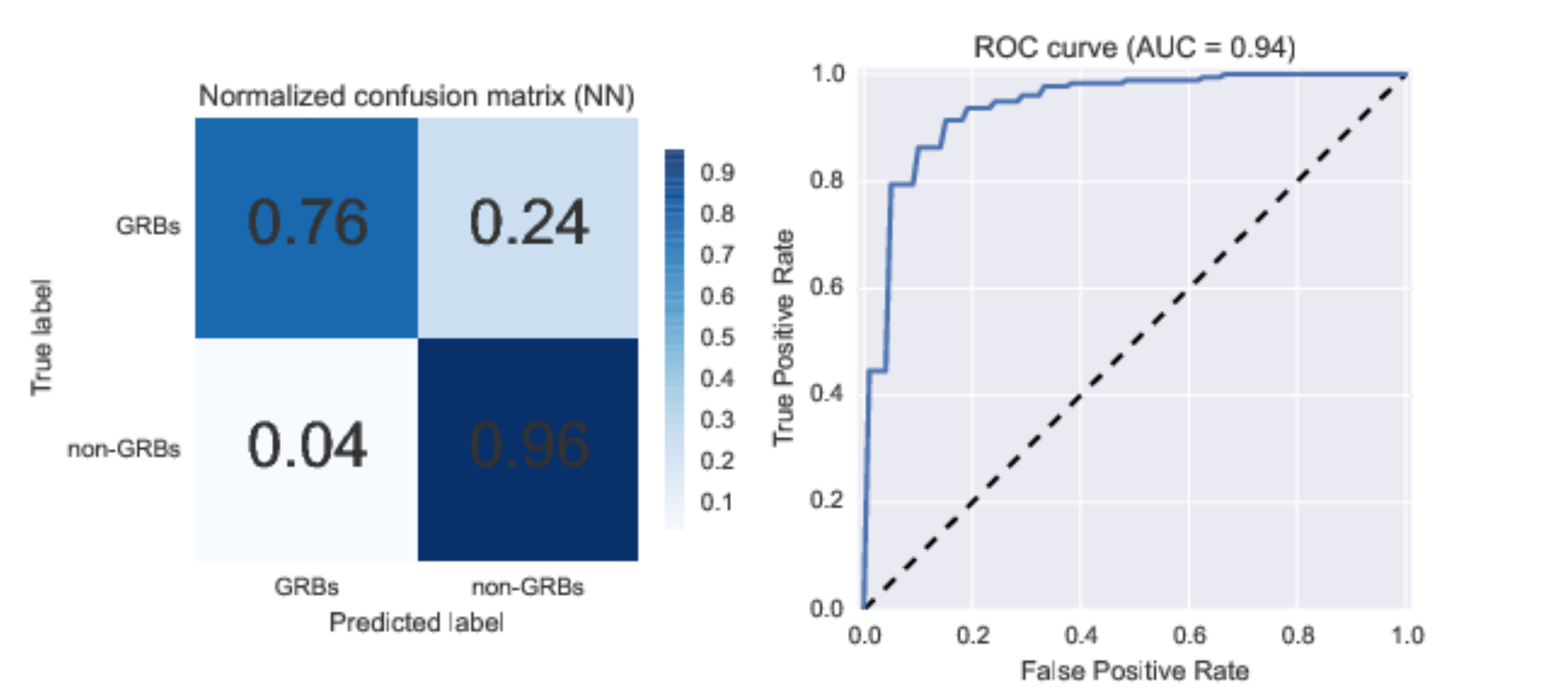}
   \caption{Confusion matrix (left) and the ROC curve (right) for the NN-classifier. The overall accuracy is 0.90.}
    \label{fig:nn-cmroc}
\end{figure}

The comparison of CMs and the ROC curves with area under ROC curve
is presented in Fig.~\ref{fig:svm-cmroc}, Fig.~\ref{fig:rf-cmroc} and Fig.~\ref{fig:nn-cmroc}.

\begin{table*}[!ht]
\caption{Summary of different scoring metrics describing the performance of the individual classifiers and the combined meta-classifier: Overall accuracy, precision, recall, f1-score, area under the ROC curve (AUC) and Matthew correlation coefficient.}
\label{table:accuracy}
\smallskip
\begin{center}
{\small
\begin{tabular}{lcccccc}  % l = left, c = centered
\hline
\noalign{\smallskip}
Classifier & Accuracy & Precision & Recall & f1-score & AUC & Matthew corr.\\
\hline
\noalign{\smallskip}
SVM & 0.90 & 0.91 & 0.95 & 0.93 & 0.95 & 0.78 \\ 
RF & 0.89 & 0.93 & 0.91 & 0.92 & 0.94 & 0.76 \\
NN & 0.90 & 0.89 & 0.96 & 0.93  & 0.94 & 0.76 \\
meta & 0.92 & 0.92 & 0.93 & 0.94 & 0.96 & 0.80 \\ 
\noalign{\smallskip}
\hline
\end{tabular}
}
\end{center}
\end{table*}

\begin{figure}[ht!]
   \centering
   \includegraphics[width=0.95\linewidth]{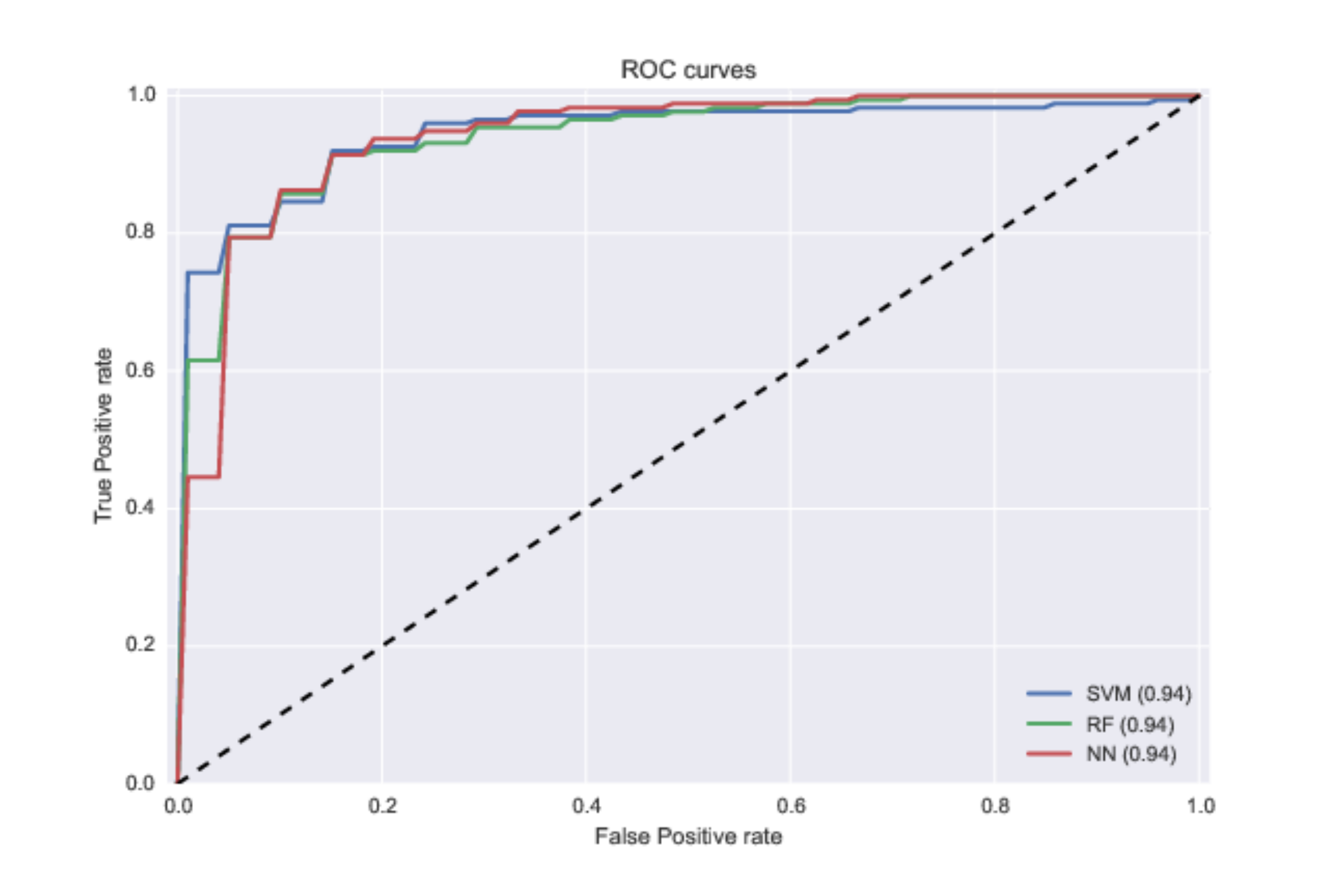}
   \caption{Classifier method comparison based on ROC curve for three different algorithms: SVM, RF and NN. Area under the ROC curve is plot. Dashed line represents the accuracy of a random classifier.}
   \label{fig:roccomp}
\end{figure}

The importance of individual features obtained from RF are summarised in the Table~\ref{table:importance}.

\begin{table}[!ht]
\caption{Feature importance deduced from Random Forest pruning.}
\label{table:importance}
\smallskip
\begin{center}
{\small
\begin{tabular}{cc}  % l = left, c = centered
\hline
\noalign{\smallskip}
Feature name & Importance [\%]\\
\hline
\noalign{\smallskip}
'g-r' & 23.8\\
'r-i' & 33.5 \\
i'-z' & 42.7 \\
\noalign{\smallskip}
\hline\
\end{tabular}
}
\end{center}
\end{table}

Each classifier uses a number of hyper-parameters that affects the overall performance of the search. Best values were find using a grid search for the minimal error in the terms of the Matthews correlation metric in Eq.~\ref{eq:metrics} over the parametric space. The results are the product of K-fold cross-validation with $K=4$. The best setups for all three classifiers are summarised in the Table~\ref{tab:hyperpar}.

\begin{table*}[!ht]
\caption{Summary of hyper-parameters used in the analysis of the complete dataset. The values were obtained by optimising the classification performance through K-fold cross-validated. The Matthew correlation coefficients were used as the scoring metrics.}
\label{tab:hyperpar}
\smallskip
\begin{center}
{\small
\begin{tabular}{lll}  % l = left, c = centered
\hline
\noalign{\smallskip}
SVM & L2 penalty (regularisation term) parameter & $C = 0.76$ \\
~   & Kernel                              & radial base function \\
~   & Kernel coefficient                  & $\gamma = 1.56$ \\
\hline
\noalign{\smallskip}
RF  & Number of trees                     & $n_{estimators} = 400$ \\
~   & Maximal depth of a tree             & $depth_{max} = 3$ \\
~   & Splitting criterion                 & Hellinger \\
~   & Maximal number of features used for split & $3$ \\
\hline
\noalign{\smallskip}
NN  & hidden layer topology                      & $(4, 4)$ \\
~   & initial learning rate               & $3.33$ \\ 
~   & L2 penalty (regularisation term) parameter & $\alpha = 0.010$ \\
~   & Activation function                 & tanh \\
\noalign{\smallskip}
\hline
\end{tabular}
}
\end{center}
\end{table*}

\subsection{Meta Classifier}
\begin{figure}[ht!]
   \centering
   \includegraphics[width=0.95\linewidth]{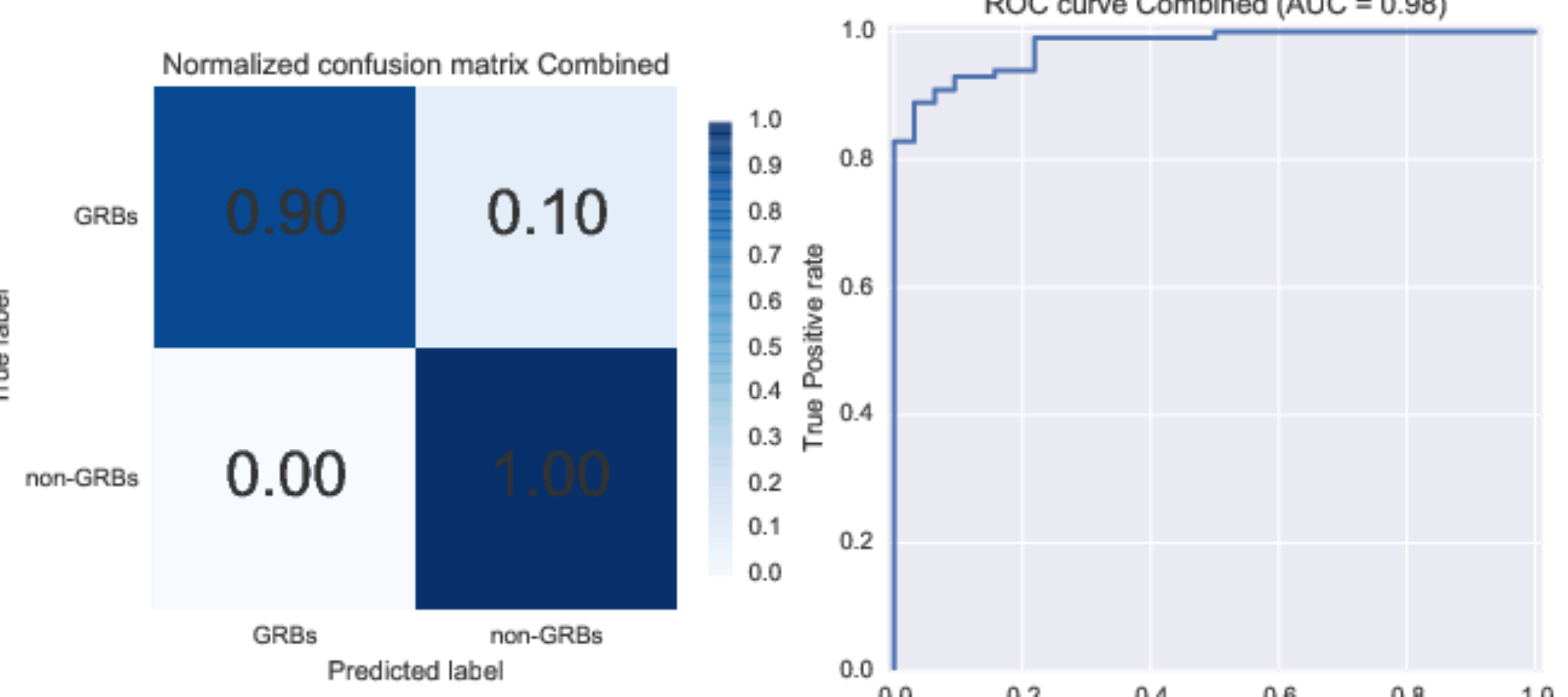}
   \caption{Confusion matrix (left) and the ROC curve (right) for the meta-classifier built from the SVM, RF and NN classifiers. The overall accuracy is 0.95.}
    \label{fig:meta-cmroc}
\end{figure}

To increase the sensitivity and specificity classification performance a meta-classifier is created by combining the powers of all three individual classifiers. The source is set to be a GRB if all three classifiers classified the object as a GRB. The scoring is in the Table~\ref{table:accuracy} and  the corresponding confusion matrix and the ROC curve is shown in Fig.~\ref{fig:meta-cmroc}.

\subsection{Redshift}
\begin{figure}[ht!]
   \centering
   \includegraphics[width=0.95\linewidth]{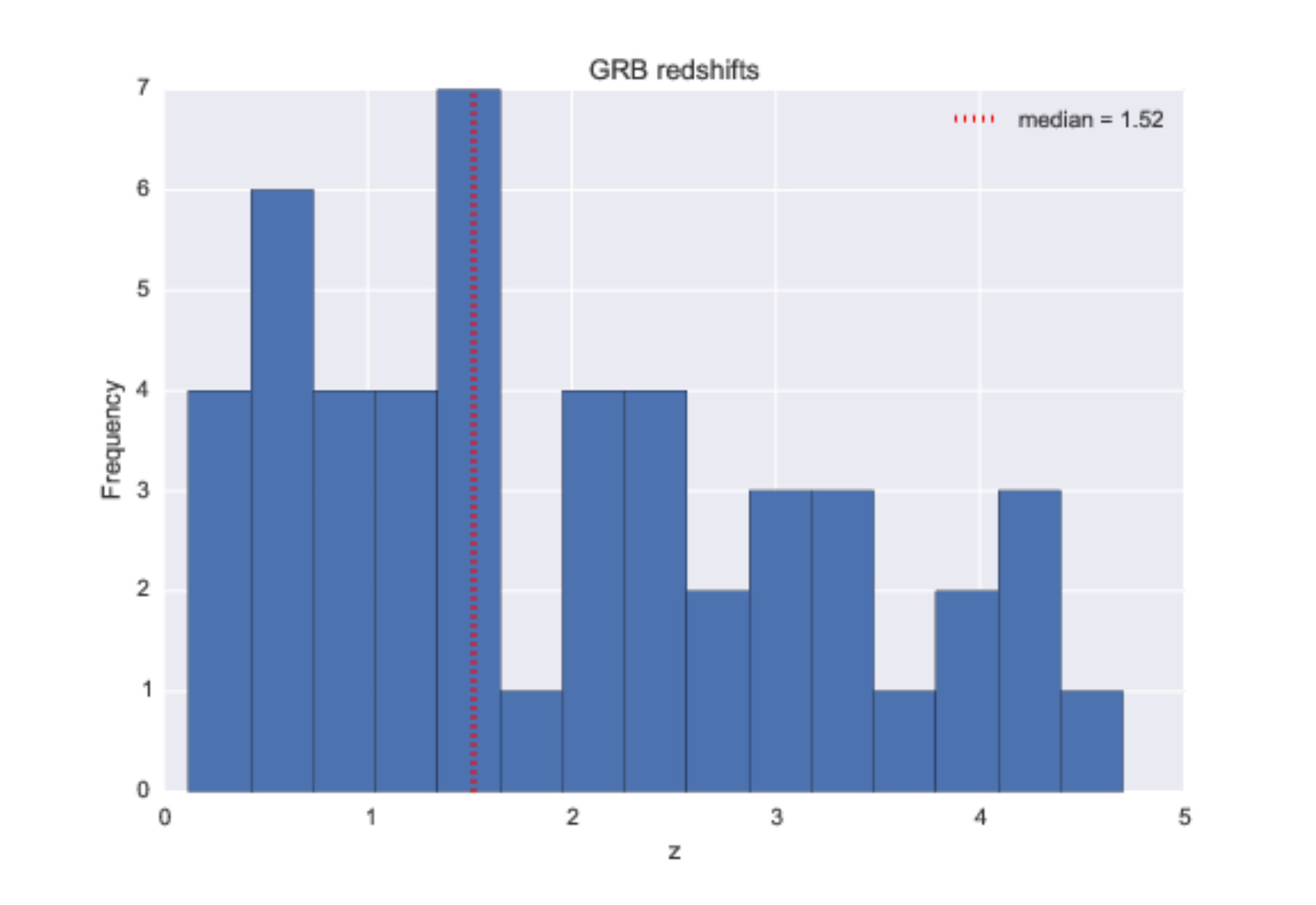}
   \caption{Redshift distribution of GROND GRB afterglows used in the analysis. 49 of 84 (58\%) GRBs have derived redshift. The median of $z_{median} = 1.52$ is marked with a red dotted line.}
    \label{fig:redshifts}
\end{figure}

58\% of the GROND GRB afterglows in the sample have measured redshift\footnote{\url{http://www.mpe.mpg.de/~jcg/grbgen.html}} $z$. The histogram of measured redshifts is plot in Fig.~\ref{fig:redshifts} with median of $z_{median} = 1.52$. To measure the classification performance on GRBs of different redshifts, the GRB sample is split into low-z and high-z samples, with respect to the median redshift $z_{median}$.

\begin{figure}[ht!]
   \centering
   \includegraphics[width=0.95\linewidth]{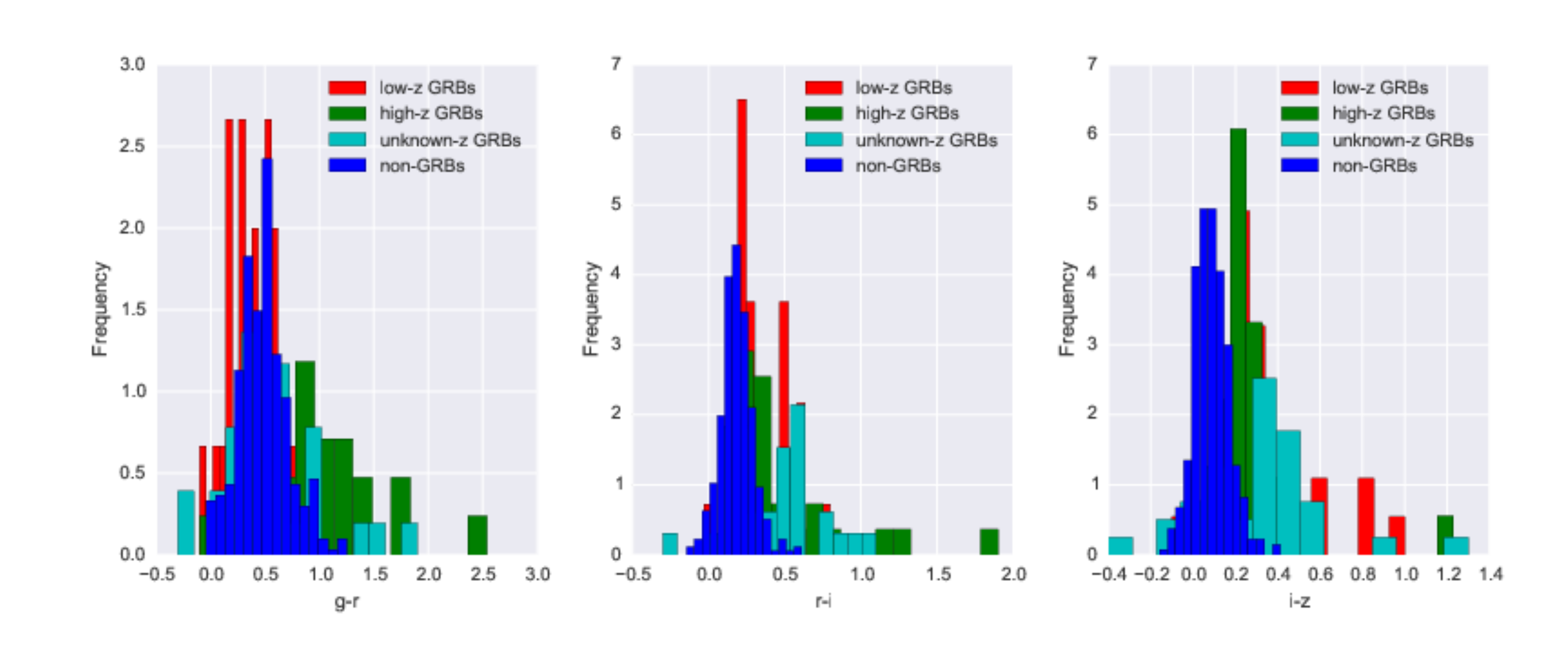}
   \caption{Stacked distribution of colour indices for the GROND GRB afterglows of different redshifts used in the analysis.}
   \label{fig:color-hist}
\end{figure}

The distribution of g'-r', r'-i', i'-z' spectral indices in each category of redshifts (low, high, unknown) is shown in Fig.~\ref{fig:color-hist}.

The GRBs are split to three groups: low-z ($z<=z_{median}$), high-z ($z>z_{median}$) if the direct or photometric redshift is known, and unknown-z if the redshift information is not provided.
Table~\ref{table:oneclassredshift} shows the fraction of correctly classified GRB afterglows in the entire dataset in each redshift group.

\begin{table}[!ht]
\caption{Classification accuracy in correctly classified GRB afterglows for afterglows with low, high and unknown redshifts. The algorithm has been trained on the entire dataset.}
\label{table:oneclassredshift}
\smallskip
\begin{center}
{\small
\begin{tabular}{lccc}  % l = left, c = centered
\hline
\noalign{\smallskip}
Classifier & Low-z & High-z & Unknown-z \\
\hline
\noalign{\smallskip}
SVM & 0.95 & 0.84 & 0.90 \\
RF &  0.92 & 0.95 & 0.95 \\
NN &  0.80 & 0.67 & 0.70 \\
\noalign{\smallskip}
\hline\
\end{tabular}
}
\end{center}
\end{table}

Alternatively, the classification performance is measured if only low-z, respectively high-z GRB subset is used to train and test the model. The results for datasets restricted to the low-z, high-z and unknown-z GRB subsets are shown in Table~\ref{table:classredshift}.
\begin{table}[!ht]
\caption{Classification accuracy in correctly classified GRB afterglows for afterglows with low, high and unknown redshifts. The algorithm has been trained on the entire dataset.}
\label{table:classredshift}
\smallskip
\begin{center}
{\small
\begin{tabular}{lcccccc}  % l = left, c = centered
\hline
\noalign{\smallskip}
Group & Accuracy & Precision & Recall & f1-score & AUC \\
\hline
\noalign{\smallskip}
\multicolumn{6}{l}{SVM} \\  
low-z & 0.93 & 0.83 & 0.80 & 0.88 & 0.96 \\ 
high-z & 0.89 & 0.89 & 0.89 & 0.89 & 0.88 \\
unknown-z & 0.92 & 0.74 & 0.84 & 0.89  & 0.90 \\
\hline
\noalign{\smallskip}
\multicolumn{6}{l}{RF} \\
low-z & 0.93 & 0.83 & 0.80 & 0.88 & 0.90 \\ 
high-z & 0.94 & 0.86 & 0.86 & 0.91 & 0.96 \\
unknown-z & 0.92 & 0.74 & 0.84 & 0.89  & 0.90 \\
\hline
\noalign{\smallskip}
\multicolumn{6}{l}{NN} \\
low-z & 0.93 & 0.83 & 0.80 & 0.88 & 0.90 \\ 
high-z & 0.94 & 0.86 & 0.86 & 0.91 & 0.96 \\
unknown-z & 0.92 & 0.74 & 0.84 & 0.89  & 0.90 \\
\hline
\end{tabular}}
\end{center}
\end{table}
The train and test error decrease with the number of data points, therefore direct comparison between the analysis on the full size data set and the low-z, respectively high-z subsets only, would be inaccurate.

\section{Discussion}
Three conditions necessary to apply machine learning should be fulfilled: 1) A pattern exists - both observational and theoretical assumptions suggest that there is a difference between GRB afterglows and other sources in the optical images, 2) the pattern is not known analytically 3) the data of reasonable quality from GROND and SDSS are available.

 All three classifiers work reasonable well, scoring at 90\% overall accuracy in the GRB class and up to 94\% recall. Since the abundance in GRB and non-GRB classes is imbalanced, Matthew correlation,  f1-score and area under the ROC curve measure the classification better. If the power of the combined meta-classifier used the classification reaches 95\% overall accuracy, 90\% recall, 0.96 AUC score and minimises the false positives. The RF algorithm gets the best scores, however, the disadvantage of RF is poor ability to extrapolate data.

The redshift distribution is not flat, therefore more low-z GRBs are expected to be hidden among the unknown-z GRBs. The difference in the classification performance between the low-z and high-z redshift sub-group is subtile, below $3\sigma$ significance.

Each model returns the probability of each data point to be a GRB afterglow. The misclassified GRB afterglows are typically very peculiar cases at the edge of prediction threshold. This also partially reflects the way the input data were obtained. The information about GRBs were taken mostly from public GCN circulars but also from selected papers on peculiar GRBs published by the GROND team. 

\begin{figure}[ht!]
   \centering
   \includegraphics[width=0.95\linewidth]{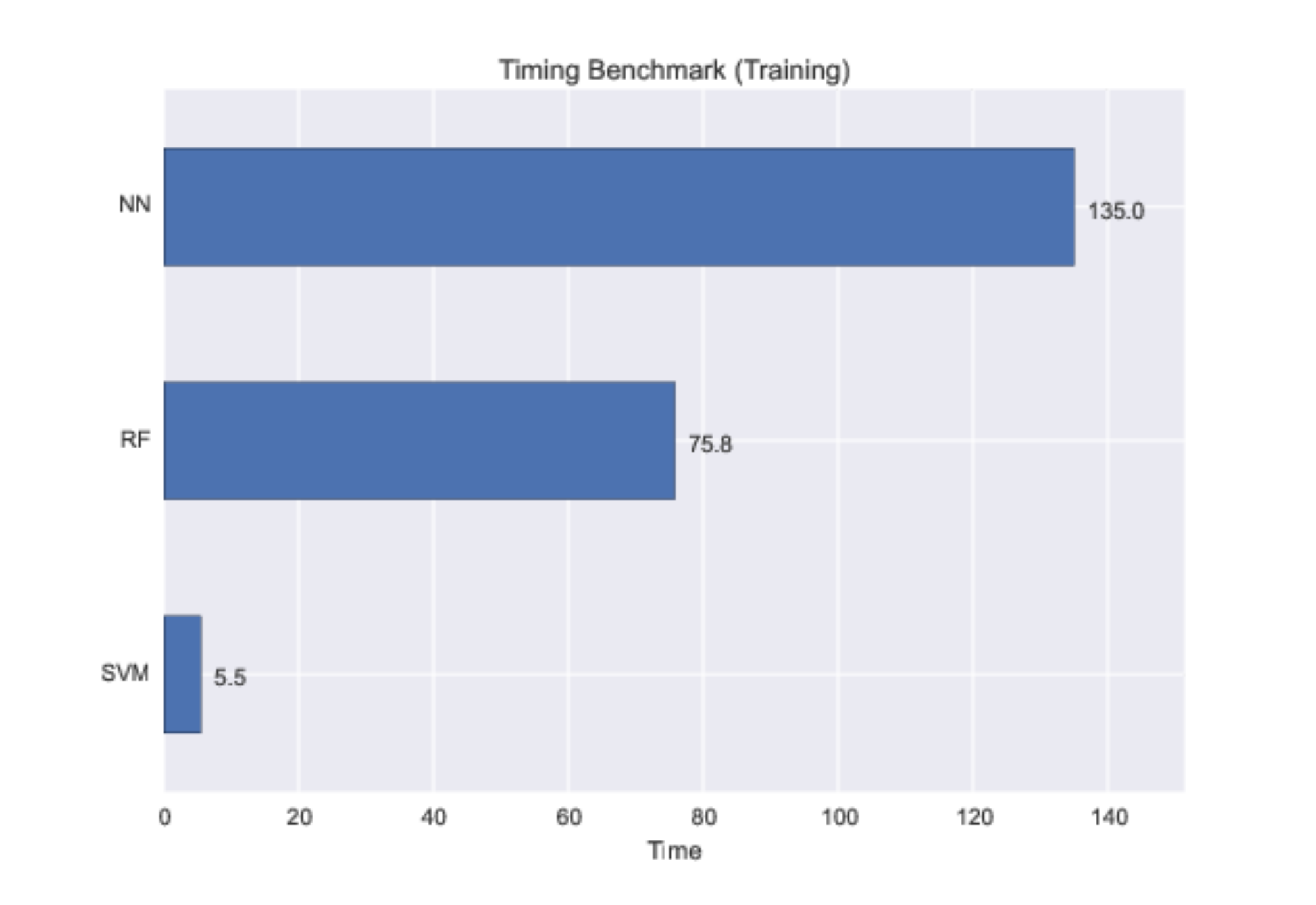}
   \caption{Benchmark of the timing analysis of SVM, RF and NN classifiers. The CPU time spent in the training phase is shown.}
   \label{fig:timing}
\end{figure}

The timing analysis of each algorithm in terms of the CPU time during the learning phase is visualised in Fig.~\ref{fig:timing}.  The advantage of SVM lies in its low CPU demand, while NN can accomplish complex non-linear rules. These scores reflect the CPU demands in the model training phase. The classification of a new GRB candidate is lightning fast once the the best classification meta-parameters are pre-computed and the classification model is fit. The complexity of the best matrix multiplication algorithm (involved in SVM and NN scales as $\mathcal{O}(k n^{2.38})$, while RF scales as $\mathcal{O}(k n \log_2 n)$ where $k$ is the number of extra layers in NN or the size of the RF forest.

This analysis has been done without using the information about the co-ordinates of the source, neither wether the source location lies within a known GRB error-box. If additional localisation information is provided, the classification accuracy increases significantly, reaching 99.99\%.

\section{Conclusion}

The analysis showed that machine learning is powerful tool for source classification that can be applied on GRB afterglows with high accuracy.
If temporal evolutionary data are missing (which is often the case in historical surveys), it has been shown that the minimal set of features including at least three colour indices g'-r', r'-i', i'-z' is sufficient to reveal $\sim 90$\% of GROND GRB afterglows.

GRB colours occupy large volume in the parametric space. Predicted GRBs can be confused with flares of non-GRB origin that can posses similar spectral properties.

The feature importance based on RF pruning  yield that the used features are of comparable order of magnitude. This means that further reduction of features would reduce dramatically the classification ability of the algorithms. On the other hand, if more relevant features were available, as suggested in Table~\ref{table:features}, even more precise classification and faster convergence is expected.

The analysis shows high importance of multi-filter observations for robotic telescope while they are in survey mode. Alternatively, three or more time-stamps in less filters and extensive contextual information should be use.

While the training phase can be CPU demanding for very crowded fields depending on the volume and variety of the dataset used for learning, the classification of a new source is lightning fast, involving simple matrix multiplication (SVM, NN) and walk through a binary tree (RF), while the matrix multiplication operation is easy compute in parallel. Also the training phase of all three algorithms can be parallelised to achieve close to real-time performance. 

Therefore, applying the classification in real-time to accomplish a rapid follow-up observation is feasible. The fast classification gives high chance not to miss an early afterglow emission with the contribution of the reverse shock and to catch the temporarily coincident $\gamma$-ray and optical emission. The exact radiation mechanism of GRBs has not been well understood yet. Studies of such correlations between the $\gamma$-rays and optical radiation emitted during the prompt phase of the burst put constraint on the GRB radiation models.

Applying the classifier would make the GRB afterglow detections possible in cases of GRBs with very large error-boxes, such is often the case of many FERMI GRBs and in the search for gravitational wave counterparts where wide-field cameras with large field of view are used.

Another application would be object classification in the existing or past sky surveys.  There exists a vast number of archival photographic plates, often with low dispersion spectra provided, e.g. First Byurakan Survey (FBS), Second Byurakan Survey (SBS) etc. The plates are of a huge and unique scientific potential, hiding many unidentified and flaring sources. The classification analysis could identify possible GRB afterglow candidates, including orphan afterglows.
Once, the true error-rate of the classification has been estimated, e.g. based on the successful follow-up observations of afterglow candidates, the true abundance of GRB afterglows can be deduced.
The orphan afterglow rate directly impacts the beaming angle and therefore the true rate and the true energy of the GRB explosions.

Statistical classification leads to detection of outliers. The outliers are peculiar cases in which the object lies at the boundary decision or at far distance from the rest of the class members in the parametric space. Pinpointing the outliers which often leads to new discoveries.

\begin{figure}[ht!]
   \centering
   \includegraphics[width=0.95\linewidth]{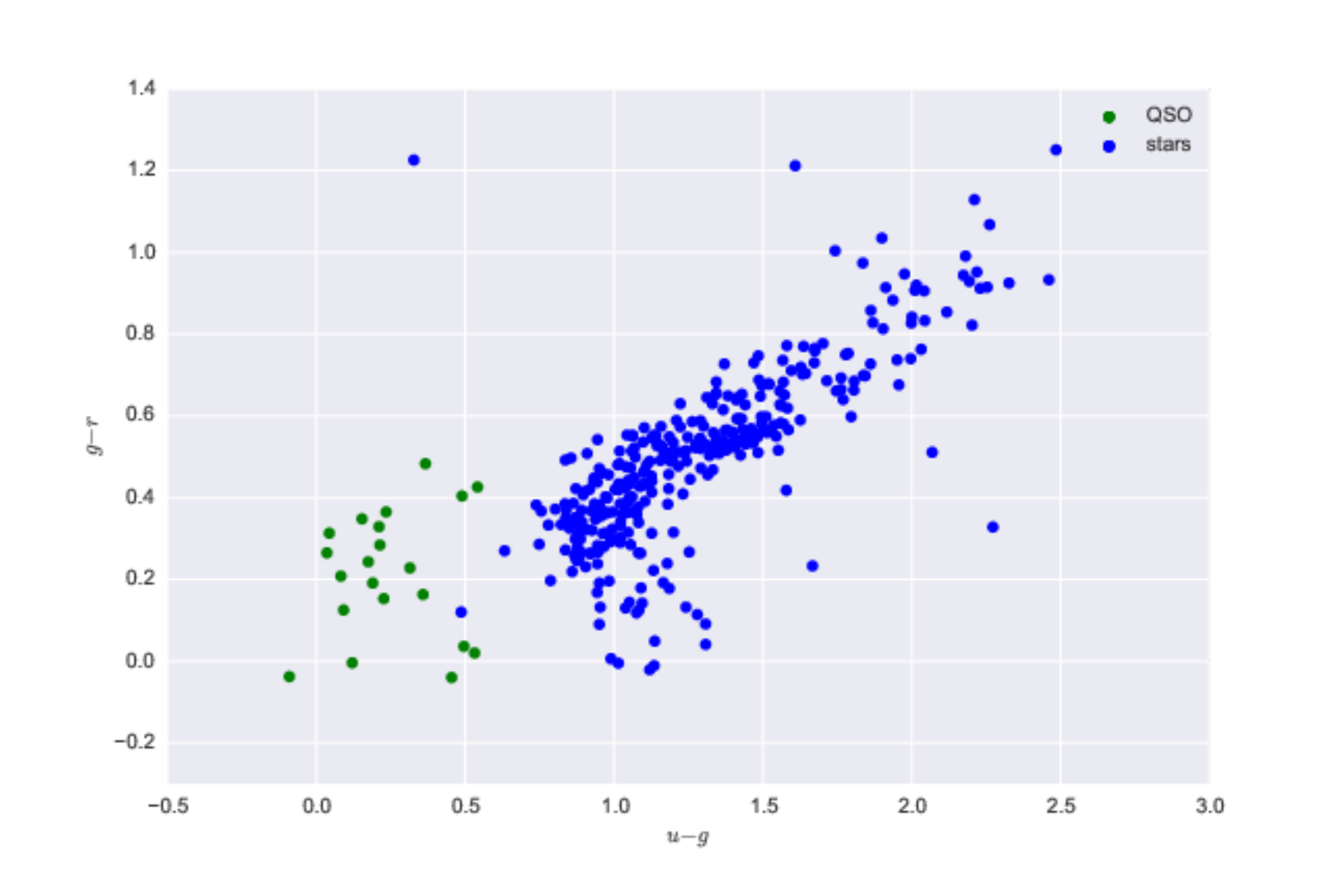}
   \caption{An example of a separation between QSO and star objects based on u'-g' and 'g-r' colour-colour diagram for the data used in this analysis.}
   \label{fig:qso-stars}
\end{figure}

The ML classification methods presented here are extensible to a broader spectrum of astrophysical objects and can be used in large surveys. Different algorithms have their perks and flaws and work best on different classification task. It has been shown that a decision tree composed of a series of binary classifiers working on different set of features yields significantly higher scores than a single \textit{multi-label} classification.
It is suggested that the classifier sensitive to GRB afterglows would be close to the top root of the decision tree.
For example, both observed quasars and GRB afterglows are believed to be blue colour objects. It has been shown that QSOs and stars are separable in the $u'-g'$ vs. $g'-r'$ colour-colour diagram based on the SDSS observations \citep{2014sdmm.book.....I}. For sake of completeness, the separation between QSOs and stars used in this analysis is clear (Fig.~\ref{fig:qso-stars}).

\acknowledgements This publication was supported by the European social fund within the framework of realising the project ,,Support of inter-sectoral mobility and quality enhancement of research teams at Czech Technical University in Prague'',
CZ.1.07 / 2.3.00 / 30.0034.
Period of the project's realisation 1.12.2012--30.9.2015.

\bibliographystyle{aa}
\bibliography{aa-topinka}
\end{document}